\documentstyle[12pt,epsfig]{article}
\begin{document}

\newcommand{\al}{\mbox{$\alpha$}}
\newcommand{\be}{\mbox{$\beta}}
\newcommand{\g}{\mbox{$\gamma$}}
\newcommand{\de}{\mbox{$\delta$}}
\newcommand{\om}{\mbox{$\omega$}}
\newcommand{\Om}{\mbox{$\Omega$}}
\newcommand{\tri}{\mbox{$\triangle$}}
\newcommand{\tr}{\mbox{$\tilde{r}$}}
\newcommand{\tom}{\mbox{$\tilde{\omega }$}}

\def\beq{\begin{equation}}
\def\eeq{\end{equation}}
\def\beqa{\begin{eqnarray}}
\def\eeqa{\end{eqnarray}}
\def\s{\sigma}   
\def\ve{\varepsilon}
\def\bl{\bar{\lambda}}
\def\da{\dagger}
\def\la{\lambda}   
\def\ti{\tilde}
\def\pr{\prime} 
\def\f{\frac}
\def\sq{\sqrt}
\def\pa{\partial}

\begin{titlepage}
\begin{flushright}   RRI--97--2;   
		     gr-qc/9701040\\
\end{flushright}
\begin{center}
   \vskip 3em 
   {\LARGE Quantum Aspects of Ergoregion Instability}
   \vskip 1.5em
   {\large Gungwon Kang\footnote{kang@rri.ernet.in} 
   \\[.5em]}
{\em Raman Research Institute, Bangalore 560 080, 
India}  \\[.7em]
\end{center}
\vskip 1em

\begin{abstract}

It has been known classically that a star with an ergoregion but no event 
horizon is unstable to the emission of scalar, electromagnetic and 
gravitational waves. This classical ergoregion instability is characterized
by complex frequency modes. We show how to canonically quantize a neutral
scalar field in the presence of such unstable modes by considering a simple 
model for a rapidly rotating star. Some of interesting results is that 
there exists a physically meaningful mode decomposition including unstable 
normal mode solutions whose representation turns out to be  
a non-Fock-like Hilbert space. A ``particle" detector model placed in 
the in-vacuum state also shows that stars with ergoregions give rise to 
a spontaneous energy radiation to spatial infinity until ergoregions 
disappear.

\vspace{1cm}
\noindent
PACS number(s): 04.40.Dg, 04.62.+v, 04.60.Ds, 97.10.Kc    
\end{abstract}
\end{titlepage}

\newpage
\section{Introduction }
\label{In}

In general relativity, inertial frames around rotating objects are dragged 
in the sense of the rotation. If the object is rotating rapidly, this 
dragging effect can be so strong that in some region no physical object 
can remain at rest relative to an inertial observer at spatial infinity. 
This region of spacetime is called an ergoregion or an ergosphere. 
For a stationary asymptotically flat spacetime, the ergosphere is the 
region in which the Killing vector field $\xi = \partial /{\partial t}$ that 
corresponds to time translations at spatial infinity fails to be timelike.
The most common example of ergoregions would be the outside of the 
event horizon of any rotating black hole. Ergoregions also arise in models 
of dense, rotating stars in which no event horizon exists inside the 
ergosurface \cite{BI,SC}. The question we want to answer in this paper
is whether or not and how the presence of ergoregions only gives rise 
to quantum instability. 

For massless field perturbations in rotating black holes, it has been shown 
that no unstable mode occurrs classically \cite{DI,Whiting}. However, 
quantum fields in this background spacetime where both horizon and 
ergoregion  are present reveal vacuum instability, the so-called 
Starobinskii-Unruh effect, which is indeed the quantum counter part of the
classical phenomena such as Penrose process for particles and 
superradiance for waves \cite{Staro,Unruh,Ford,AM1}. On the other hand, 
spacetimes with ergoregions but without horizons such as rapidly rotating 
stars are known to be unstable to scalar, electromagnetic and 
gravitational perturbations \cite{Fried,CS}. As explained in 
Refs.~\cite{Vilen,Fried,CS,AM1}, such classical instability, so-called 
ergoregion instability, can be understood heuristically by looking at 
a spherical wave packet in a ``superradiant" mode incident to the 
ergoregion. The reflected wave will be amplified and carry positive 
energy up to spatial infinity. The transmitted part within the ergoregion 
is also amplified but now carries negative energy with respect to an
observer at infinity. This wave will pass through the center of the 
rotating object and get back to the ergosurface, again giving transmission
as well as reflection there with energies amplified, repectively. 
This process will repeat as long as the ergoregion remains, resulting in 
presumably  ``exponential" radiation of positive energy to infinity 
and accumulation of negative energy within the ergoregion in such a way that
the total energy is conserved. It turns out that this instability is
characterized by complex frequency modes in normal mode solutions of
classical fields. 

The quantum counterpart of such ergoregion instability has been studied 
by Ashtekar and Magnon \cite{AM2,AM1}, and Dray, Kulkarni and Manogue 
\cite{DKM} in algebraic approach, and by Matacz, Davies and Ottewill in 
canonical approach \cite{MDO} of quantum fields. Ashtekar and Magnon have 
given  a general argument in Ref.~\cite{AM2} that in a rotating spacetime
with an ergoregion but without a horizon there does not exists a natural
complex structure which gives no ``mixing of positive and negative 
frequencies" under time evolution, and hence no spontaneous creation of 
particles. In other words, no matter which complex structure is chosen,  
the resulting quantum field theory should predict a spontaneous particle
creation. However, their algebraic approach is somewhat qualitative and
it has not been shown what the detailed quantitative pictures of this
quantum instability are. On the other hand, Dray, Kulkarni and Manogue 
have claimed that there exists a natural complex structure if we quantize 
the system with respect to ZAMO observers \cite{ZAMO} rather than with 
respect to Killing observers. 

In the canonical quantization approach, 
Matacz, Davies and Ottewill have recently investigated whether quantum 
vacuum instability occurrs near rapidly rotating stars. They have 
considered a simple spacetime model where the outside of a star is 
described by the Kerr metric, and classical Klein-Gordon fields satisfy
a mirror boundary condition (i.e., $\phi (x) = 0$) near the surface of 
the rotating star. They found that the quantum instability, the 
Starobinskii-Unruh effect, is {\it absent} provided that only real 
frequencies occur in normal mode solutions of the Klein-Gordon equation. 
As mentioned above, however, all asymptotically flat stationary 
spacetimes with ergoregions but without horizons are unstable to scalar 
wave perturbations. Consequently, complex frequency modes should exist 
in such a background spacetime. 
Therefore, in order to conclude whether or not the
quantum instability occurrs near stars with ergoregions, one must include
such unstable complex frequency modes as well and needs to understand 
their physical roles
in the quantization procedure. 

After the discovery of the occurrence of complex frequency modes for a 
charged scalar field interacting with some strong electrostatic potential, 
the so-called Schiff-Snyder-Weinberg effect \cite{SSW}, the quantization 
of fields including such inatability modes has been studied by several
authors, all in the Minkowski flat spacetime; a charged scalar field
interacting with strong stationary scalar potential \cite{SS}, a neutral 
scalar field with tachyonic mass \cite{Schroer}, a charged scalar field 
in an electrostatic potential \cite{Fulling}, and a neutral scalar field 
with time-varying mass or interacting with an external square-well 
potential \cite{Kang}. 

In this paper we carry out, for the first time, 
this quantization procedure for a massless scalar field in a certain 
model of curved spacetime with an ergoregion but without a horizon. 
We find that it is possible to quantize the system, but mode operators for 
complex frequencies satisfy unusual commutation relations and do not 
admit a usual Fock-like representation or a ground state as in other models
in Minkowski flat spacetime. Consequently, the results imply that 
a rotating star with an ergoregion gives a spontaneous energy radiation 
to spatial infinity until the ergoregion disappears. This quantum 
ergoregion instability is very much analogous to a laser amplification. 

In Sec.~\ref{CF}, we define the inner product for field solutions and show
its peculiar properties in the presence of complex frequency modes. 
We also explain the spacetime model to be considered and construct a complete
set of normal mode solutions including instability modes due to the 
presence of ergoregion. In Sec.~\ref{Q}, the canonical quantization is
carried out, which turns out to give a non-Fock-like representation for
the quantum field. In Sec.~\ref{QI}, a ``particle" detector model is
used to show the spontaneous energy radiation to infinity related to
the appearance of the ergoregion. Some open questions and possible 
applications of our results to other physical systems are discussed in 
Sec.~\ref{D}.

\section{Classical field}
\label{CF}

We consider a system of a massless real scalar field $\phi (x)$ minimally 
coupled to gravitational fields as follows,
\beq
{\rm L} =\int d^3x \sqrt{-g}{\cal L}  
        =-\frac{1}{2}\int d^3x\sq {-g} g^{\mu \nu}\pa_{\mu}\phi 
         \pa_{\nu}\phi  , 
\eeq 
where the integration is performed on a $x^0=t={\rm const.}$
spacelike hypersurface and $g^{\mu \nu}$
is the metric of an arbitrary background spacetime which will be specified 
below. The conjugate momentum $\pi$ to the field $\phi$ is defined 
by
\beq
\pi (x) = \frac{\partial {\cal L}}{\partial (\partial \phi /{\partial x^0})}
    = -g^{0\nu}\pa_{\nu}\phi = -\nabla^0\phi ,
\eeq
and the Hamiltonian is
\beqa
{\rm H} &=& \int_{x^0={\rm const.}}d^3x\sq{-g} (\pi \nabla_0\phi -{\cal L})
            \nonumber  \\
        &=& \frac{1}{2}\int d^3x\sq{-g}(-g^{00}\nabla_0\phi \nabla_0\phi 
             +g^{ij}\nabla_i\phi \nabla_j\phi ). 
\label{Ham}
\eeqa
The Klein-Gordon equation
\beq
\Box \phi = \frac{1}{\sq{-g}}\pa_{\mu}(\sq{-g}g^{\mu \nu}\pa_{\nu}\phi )=0
\label{KG}
\eeq
is equivalent to Hamiltonian equations given by 
\beq
\hat{\rm H} \Phi = i \f{\pa \Phi}{\pa x^0},
\label{FEQ}
\eeq
where $\Phi =\left(\matrix{\phi \cr \pi \cr}\right)$ is the two-component
field, and $\pa \phi /{\pa x^0} = \pa {\cal H}/{\pa \pi}$ and 
$\pa \pi /{\pa x^0} = -\pa {\cal H}/{\pa \phi}$. As shown in 
Ref.~\cite{Unruh}, the operator $\hat{\rm H}$ defined via Eq.~(\ref{FEQ})  
is Hermitian with 
respect to the following inner product 
\beqa
<\! \phi_1\, ,\, \phi_2\! > &=& <\! \Phi_1\, ,\, \Phi_2\! > \nonumber \\
 &=& \frac{i}{2}\int d^3x\sq{-g}\, (\phi_1^{\ast}\pi_2-\pi_1^{\ast}\phi_2) 
     \nonumber  \\
 &=& \frac{i}{2}\int \, \phi_1^{\ast}\! \stackrel{\leftrightarrow }
     {\partial_{\mu}}\! \phi_2 \,\, d\Sigma^{\mu}. 
\label{Inner}
\eeqa
In other words,
\beq
<\! \Phi_1\, ,\, \hat{\rm H}\Phi_2\! >=<\! \hat{\rm H}\Phi_1\, ,\, \Phi_2\! >
\label{Herm}
\eeq
for any given solutions $\phi_1$ and $\phi_2$ of Eq.~(\ref{FEQ}) or,
equivalently, Eq.~(\ref{KG}), satisfying suitable conditions on the 
spatial boundary. Consequently, the inner product 
$<\! \phi_1\, ,\, \phi_2\! >$ is independent of the ``time" $x^0=t$ 
at which the spatial integration is performed. 

We now assume that the background spacetime possesses a Killing vector 
field $\xi =\pa_t$ in some regions, for instance, in the early and 
late stages of its evolution. Then normal mode solutions can be defined by
\beq
{\cal L}_\xi \phi = -i \om \phi .    \nonumber  
\eeq
Here ${\cal L}_\xi$ is the Lie-derivative along a Killing vector $\xi$.
Thus the time dependence of normal mode solutions is $\sim e^{-i\om t}$.
Since ${\cal L}_\xi \phi = \pa \phi /{\pa x^0}$, normal mode solutions
are indeed eigenfunctions of the operator $\hat{\rm H}$
\beq
\hat{\rm H} \Phi = \om \Phi .    \nonumber
\eeq
From the Hermiticity of $\hat{\rm H}$ in Eq.~(\ref{Herm}), one obtains 
\beq
(\om_2-\om_1^{\ast})<\! \phi_1\, ,\, \phi_2\! > = 0
\label{OrthoG}
\eeq
for any given two normal mode solutions $\phi_1$ and $\phi_2$. Thus the inner
product is zero unless $\om_2=\om_1^{\ast}$. Since our inner product defined 
in  Eq.~(\ref{Inner}) is not positive definite in general, the normal
mode frequency $\om$ is not necessarily always real. In the remote past
where the spacetime is almost flat, the inner product is positive
definite and hence $\om$ is real. In the far future where an ergoregion 
arises, however, it is possible that there exist bounded solutions with
complex frequencies as shown in Refs. \cite{Vilen,CS,Fried} for certain 
cases of spacetime. Then, from Eq.~(\ref{OrthoG}), the norm of such complex 
frequency modes should be zero. 
Therefore, our inner product is not positive definite in the late
epoch of star evolution. In other words, we are confronted with the 
problem of quantizing fields in an indefinite inner product space. 

If one defines the classical energy associated with a solution $\phi$ as 
\beq
{\rm E} = <\! \Phi \, ,\, \hat{\rm H}\Phi \! >,
\label{CE}
\eeq
complex frequency modes give vanishing classical energy due to null norm, 
${\rm E}_{\om}=\om <\! \phi \, ,\, \phi \! >=0$, whereas real frequency 
modes give ${\rm E}_{\om}= \pm \om $ after suitable normalization of fields.
It happens probably because the negative energy spread over within the
ergoregion exactly cancels the positive energy outside. As shall explicitly 
be shown below, it should be pointed out 
that, by linearly combining complex frequency mode solutions, one can 
construct a solution whose energy defined in Eq.~(\ref{CE}) or, equivalently,
in Eq.~(\ref{Ham}) is {\it negative}. From Eq.~(\ref{Ham}) 
one can check that, although the Hamiltonian is positive definite in cases 
that spacetimes are almost flat in the past infinity, it could be negative 
for certain solutions if ergoregions appear in the future infinity. 
The above mentioned properties related to instability modes are also 
satisfied for other models studied 
in the flat Minkowski spacetime in Refs.~\cite{SS,Schroer,Fulling,Kang}. 

We now specify the background spacetime in more detail. We are interested in 
a dynamically evolving spacetime which starts from an almost flat spacetime
in the past infinity and ends up to a stationary rotating star with an
ergoregion in the future infinity. As far as we know, however, there is 
no known analytic solution representing such spacetime.
First of all, we should point out that there is no 
equivalent to Birkhoff's theorem for an axially symmetric rotating star. 
So the outside region of axially symmetric rotating stars may not be 
described by the Kerr metric and presumably depends on the details of 
the star inside. There is a spacetime solution found by Lindblom and
Brill \cite{LB} for the case of the free-fall collapse of a rotating dust 
shell where an ergoregion develops at the late epoch of its collapse. 
Unfortunately, however, the solution, which is based on the first order
approximation in the angular velocity of the shell, ceases to be valid
near this stage. For our purpose of showing quantum instability of 
ergoregions, however, it is sufficient to consider any spacetime model 
which possesses an ergoregion at the late stage of its evolution as
will be shown below.

We assume that our background spacetime is described by the Minkowski 
flat metric in the past infinity and by the Kerr metric with mirror 
boundary condition on the field $\phi$, which is used in 
Refs.~\cite{MDO,Vilen}, in the future infinity. Instead of considering 
the detailed dynamics inside the star, we simply assume that all classical
solutions $\phi$ of Eq.~(\ref{KG}) vanish on the surface of 
some sphere inside the 
ergoregion, e.g., a totally reflecting mirror boundary. The quantization 
of the field in the past infinity will be straightforward; it will have 
a Fock representation with a vacuum state $|0\! >_{\!\! \rm in}$. 
To carry out the canonical quantization in the future infinity, let us 
first construct normal mode solutions of Eq.~(\ref{KG}). 

In Boyer-Linquist coordinates, the Kerr metric outside the star is given by
\beqa
ds^2 &=& -\f{\tri \Sigma}{(r^2+a^2)^2-\tri a^2\sin^2\theta }dt^2 
      +\f{\Sigma}{\tri}dr^2 +\Sigma d\theta^2    \nonumber   \\
     & & +\f{(r^2+a^2)^2-\tri a^2\sin^2\theta }{\Sigma}\sin^2\theta 
         (d\varphi -\Omega dt)^2,
\eeqa
where $\tri =r^2-2Mr+a^2$, $\Sigma =r^2+a^2\cos^2\theta $, 
$\Om =2aMr/[(r^2+a^2)^2-\tri \sin^2 \theta ]$, and $M$ and $a$ 
are the total mass and the angular momentum per unit mass of the star,
respectively. This metric has a rotational Killing vector field
$\xi_{\varphi}=\pa_{\varphi}$ commuting with $\xi_t=\pa_t$. 
Thus one can simutaneously define angular eigenmodes by
\beq
{\cal L}_{\xi_{\varphi}}\phi = im\phi ,
\eeq
where $m$ is an integer. As is well known, the Klein-Gordon equation is 
separable \cite{SEP} and admits a complete set of normal mode solutions 
of the form
\beq
\phi (x) =\f{R(r)}{\sq{r^2+a^2}}S(\theta )e^{-i\om t+im\varphi }.
\eeq
Here $S(\theta )$ is the oblate spheroidal harmonics with eigenvalue 
$\lambda$ satisfying 
\beq
[\f{1}{\sin \theta}\f{d}{d\theta}\sin \theta \f{d}{d\theta}+
(\om^2a^2\cos^2\theta -\f{m^2}{\sin^2\theta})]S(\theta ) 
= -\lambda S(\theta ).
\label{KGA}
\eeq
$\lambda$ is a separation constant, a function of $\om$, $m$ and some
integer $l$ with $|m| \leq l$, and determined by requiring $S(\theta )$ 
to be regular at $\theta =0$, $\pi$. The radial function $R(r)$ 
satisfies
\beq
[\tri \f{d}{dr}\tri \f{d}{dr}+(\om (r^2+a^2)-ma)^2-(\om a(\om a-2m)+\lambda )
\tri ]\f{R(r)}{\sq{r^2+a^2}} =0.
\label{KGR1}
\eeq
Defining a ``generalized" tortoise coordinate $\ti{r}$ by
$d\ti{r}/{dr}=(r^2+a^2)/\tri $, Eq.~(\ref{KGR1}) becomes 
\beq
\f{d^2R}{d\ti{r}^2} - V_{\om lm}(\ti{r})R = 0.
\label{KGR2}
\eeq
From the mirror boundary condition, the radial function vanishes at some 
$r=r_0$ (accordingly, $\ti{r}=\ti{r}_0$) inside the ergoregion; 
\beq
R(r_0)=0.
\label{MBC}
\eeq
For simplicity, we assume that $r_0$ is very near the ``horizon" radius
$r=r_{\rm H}=M+\sq{M^2+a^2}$. That is, $\ti{r}_0 \sim -\infty $. 
We also require that the field is not singular at spatial infinity, 
$\ti{r} \sim \infty $. 
The asymptotic behavior of the effective potential $V_{\om lm}$
induced through the interaction with gravitational fields is as follows
\cite{footnote1},
\beq
V_{\omega lm}(r) \sim \left\{\begin{array}{ll}
                               -(\omega -m\Omega_H)^2  & 
\mbox{as $\tilde{r} \rightarrow \tilde{r}_0$,}    \\ 
                           -\omega^2     \qquad   &
\mbox{as $\tilde{r} \rightarrow \infty $,} 
			     \end{array}
                      \right. 
\label{Poten}
\eeq
where $\Om_{\rm H}=a/{2Mr_{\rm H}}$. We have $V_{\om lm} = \infty $ 
at $\ti{r}\sim \ti{r}_0$, corresponding to the mirror boundary condition. 
Since $\om$ could be complex in our model,
$V_{\omega lm}(r)$ is a complex potential in general. For real $\om$,
we see that, between two asymptotic regions, e.g., $\ti{r}\sim \infty $ and
$\ti{r}\sim \ti{r}_0$, there exists a potential barrier which grows as 
$l$ increases. Note that the left asymptotic value 
$-\ti{\om}^2=-(\om -m\Om_{\rm H})^2$ varies from $0$ to $-\infty$ as 
$m\Om_{\rm H}$ changes. In particular, a deep potential well is formed 
for a large value of $m\Om_{\rm H}$ which indeed leads to superradiance 
when $\ti{\om}=\om -m\Om_{\rm H} < 0$ in the case of the kerr black hole. 
Thus, in the asymptotic regions, the form of the radial solution $R(r)$ 
will be
\beq
R(r) \sim \left\{\begin{array}{ll}
                 e^{\pm i\ti{\om}\ti{r}}    & 
\mbox{as $\ti{r} \rightarrow \ti{r}_0$,}    \\ 
                 e^{\pm i\om \ti{r}}   \qquad   &
\mbox{as $\ti{r} \rightarrow \infty $.} 
			     \end{array}
                      \right. 
\eeq

Let us now consider normal mode solutions $u_{\om lm}(r)$ to Eq.~(\ref{KGR2})
whose asymptotic forms are
\beq
u_{\omega lm}(r) \sim \left\{\begin{array}{ll}
                 B_{\om lm}(e^{i\ti{\om}\ti{r}} 
                 +A_{\om lm}e^{-i\ti{\om}\ti{r}}) & 
\mbox{as $\ti{r} \rightarrow \ti{r}_0$,}    \\ 
                 e^{-i\om \ti{r}} + C_{\om lm}e^{i\om \ti{r}}  &               
\mbox{as $\ti{r} \rightarrow \infty $.} 
			     \end{array}
                      \right.
\label{ASolR}
\eeq
From the boundary condition that they vanish at the mirror surface at
$\tr =\tr_0$, 
\beq
A_{\om lm}=-e^{2i\tom \tr_0}.
\eeq
That is, as $\tr \rightarrow \tr_0$,
\beq  
u_{\om lm}(r) \sim  B_{\om lm}e^{i\tom \tr_0}(e^{i\tom (\tr -\tr_0)} - 
                     e^{-i\tom (\tr -\tr_0)})  
             = 2iB_{\om lm}e^{i\tom \tr_0}\sin \tom (\tr -\tr_0).
\eeq  
If $\om$ is complex, $u_{\om lm}$ becomes exponentially divergent at spatial 
infinity and so we exclude this class of solutions from our construction. 
Thus $u_{\om lm}$ represents real frequency normal mode solutions. 
Now the Wronskian relations from  Eq.~(\ref{KGR2}) with the mirror boundary 
condition give
\beq
|A_{\om lm}| = |C_{\om lm}| = 1.  \nonumber
\eeq
Therefore, $u_{\om lm}(r)$ is a stationary wave without any net ingoing or 
outgoing flux with respect to ZAMO observers \cite{ZAMO}. Here $\om$
is any continuous real number. In fact, this class of real frequency normal 
mode solutions is equivalent to the set considered in Ref.~\cite{MDO} 
as a complete basis. 

As mentioned above, however, there exists another class of normal mode 
solutions with complex frequencies which describe unstable modes in the 
presence of ergoregions. This class of solutions has not been included 
in the quantization procedure in Ref.~\cite{MDO}. Let $v_{\om lm}(r)$ be 
normal mode solutions in such class whose asymptotic behaviors are   
\beq
v_{\om lm}(r) \sim \left\{\begin{array}{ll}
                 e^{i\ti{\om}\ti{r}} +R_{\om lm}e^{-i\tom \tr}  & 
\mbox{as $\ti{r} \rightarrow \ti{r}_0$,}    \\ 
                 T_{\om lm}e^{i\om \ti{r}}   \qquad   &
\mbox{as $\ti{r} \rightarrow \infty $.} 
			     \end{array}
                      \right. 
\label{ASolC}
\eeq
The mirror boundary condition is satisfied if
\beq
R_{\om lm}=-e^{2i\tom \tr_0}=-e^{2i\tom^{\rm R}\tr_0}\cdot 
	    e^{-2\tom^{\rm I}\tr_0},
\eeq
where $\tom =\tom^{\rm R}+i\tom^{\rm I}$. Thus, as $\tr \rightarrow 
\tr_0$, $v_{\om lm} \sim \sin \tom (\tr -\tr_0)$ again. And
\beq
\om^{\rm I}=\tom^{\rm I}=-\ln |R_{\om lm}|/{2\tr_0}.
\label{Imaginary}
\eeq
Note that, since the potential 
$V_{\om lm}(r)$ in Eq.~(\ref{KGR2}) is complex, the Wronskian relation
does not neccessarily give $|R_{\om lm}|=1$. 
If $|R_{\om lm}| > 1$, $\om^{\rm I} > 0$ and so $v_{\om lm}(r) \sim 
e^{-\om^{\rm I}\tr}$ as $\tr \rightarrow \infty$ and is
regular at spatial infinity. From the time-dependence
of this solution, i.e., $\sim e^{-i\om t}=e^{-i\om^{\rm R}t}\cdot 
e^{\om^{\rm I}t}$, we also notice that it represents an outgoing mode 
which is exponentially amplifying in time but is exponentially 
decreasing as $\tr \rightarrow \infty$. By making a wave packet, 
as suggested in Ref.~\cite{Vilen}, we may regard this solution as an 
outgoing wave packet with $\tom^{\rm R}=\om^{\rm R}-m\Om_{\rm H} < 0$
starting from near the mirror surface, which will bounce back and forth
within the ergoregion, and a part of which is repeatedly transmitted
to infinity, resulting in exponential amplification in time in the 
inside as well as in the outside of the ergosurface. If $|R_{\om lm}| < 1$,
$\om^{\rm I} < 0$ and so this solution corresponds to an outgoing 
decaying mode in time. However, since its radial behavior becomes 
singular at spatial infinity, we do not include this mode 
in our construction of normal mode solutions.  

For any given solution
\beq
\phi_{\om lm}(x) =\phi_{\om lm}(r, \theta )e^{-i\om t+im\varphi }
=\f{v_{\om lm}(r)}{\sq{r^2+a^2}}S_{\om lm}(\theta )e^{-i\om t+im\varphi}
\sim e^{\om^{\rm I}t}
\eeq
with $\om^{\rm I}>0$, we find that there are three linearly independent 
solutions;
\beqa
\phi^{\ast}_{\om lm}(x) &=& \f{v^{\ast}_{\om lm}(r)}{\sq{r^2+a^2}}
	S^{\ast}_{\om lm}(\theta )e^{i\om^{\ast}t-im\varphi }
	\sim e^{\om^{\rm I}t},   \nonumber     \\
\phi_{\om^{\ast}lm}(x) &=& \f{v^{\ast}_{\om lm}(r)}{\sq{r^2+a^2}}
	S^{\ast}_{\om lm}(\theta )e^{-i\om^{\ast}t+im\varphi }
	\sim e^{-\om^{\rm I}t},   \nonumber     \\	
\phi^{\ast}_{\om^{\ast}lm}(x) &=& \f{v_{\om lm}(r)}{\sq{r^2+a^2}}
	S_{\om lm}(\theta )e^{i\om t-im\varphi }
	\sim e^{-\om^{\rm I}t}.   
\eeqa
$\phi^{\ast}_{\om lm}$ is simply the complex conjugation of $\phi_{\om lm}$.
$\phi_{\om^{\ast}lm}$ is obtained simply by taking complex conjugations
of Eq.~(\ref{KGA}) and Eq.~(\ref{KGR1}). Thus we let 
$\phi_{\om^{\ast}lm}(r, \theta ) =\phi^{\ast}_{\om lm}(r, \theta )$. 
This mode represents an exponentially dacaying wave in time which 
originates at infinity. In other words, this mode is the same as 
$\phi_{\om lm}(x)$ but backward in time. $\phi^{\ast}_{\om^{\ast}lm}(x)$
is simply the complex conjugate of $\phi_{\om^{\ast}lm}(x)$. 
These linearly independent modes can be denoted by indices such as
$(\om ,l,m),\,\, (-\om^{\ast},l,-m),\,\, (\om^{\ast},l,m)$, and $(-\om,l,-m)$, 
respectively. 

For these non-stationary modes, $\om$ is discrete complex 
numbers which are determined by the details of the potential and the
boundary condition. In fact, by finding poles of the scattering 
amplitude for a more realistic model of rotating stars, Comins and 
Schutz \cite{CS} have shown that the imaginary part of the complex 
frequency for a purely outgoing mode is discrete, positive, 
and propotional to $e^{-2\beta m}$, where $\beta$ is of order unit. 
For our model, it also can be shown, from Eq.~(\ref{KGA}) and 
Eq.~(\ref{KGR1}), that complex eigenfrequencies are confined 
to a bounded region as follows \cite{DI},
\beq
 0 \leq m\om^{\rm R}, \qquad  0 \leq |\om^{\rm R}| \leq |m|\Om_{\rm H}
 \f{r_{\rm H}}{r_0},    
\eeq
and
\beq
(\om^{\rm I})^2 \leq (1+a^4/{r_0^4})(\om^{\rm R})^2+2mM(m+2a\om^{\rm R})
/{r_0^3}   
\eeq
for given $m$ and $a$. 

Now let us consider the norms of these mode solutions constructed above. 
From our definition of the inner product in Eq.~(\ref{Inner}), we find
\beqa
<\! \phi_{\om lm}\, ,\, \phi_{\om^{\pr}l^{\pr}m^{\pr}}\! > 
 &=& \frac{i}{2}\int \, \phi_{\om lm}^{\ast}(
     \stackrel{\leftrightarrow }{\partial_t} +
     \Om \stackrel{\leftrightarrow }{\partial_{\varphi}})
     \phi_{\om^{\pr}l^{\pr}m^{\pr}} N^{-1}d\Sigma ,  \nonumber  \\
 &=& \frac{1}{2}\int \, [(\om^{\pr}+\om^{\ast})-\Om (m'+m)] 
     \phi_{\om lm}^{\ast}\phi_{\om^{\pr}l^{\pr}m^{\pr}} N^{-1}d\Sigma ,
\eeqa
where we have used $d\Sigma^{\mu}=n^{\mu}d\Sigma$, 
$\,\, n^{\mu}=N^{-1}(\pa_t+\Om \pa_{\varphi})^{\mu}$, $\,\, \Om (r, \theta )=
-\pa_t\cdot \pa_{\varphi}/{\pa_{\varphi}\cdot \pa_{\varphi}}= 
-g_{t\varphi}/g_{\varphi \varphi}$, and $N=[-(\pa_t+\Om \pa_{\varphi})
\cdot (\pa_t+\Om \pa_{\varphi})]^{-1/2}=(-g^{tt})^{-1/2}$. Thus,
\beq
<\! u_{\om lm}\, ,\, u_{\om lm}\! > = \int (\om -m\Om )|u_{\om lm}(x)|^2
N^{-1}d\Sigma 
\eeq
for real frequency modes with $\om > 0$ 
\beq
u_{\om lm}(x) = \f{u_{\om lm}(r)}{\sq{r^2+a^2}}S_{\om lm}(\theta )
		e^{-i\om t+im\varphi}.
\eeq
Since $\Om (r,\, \theta ) \leq \Om_{\rm H}$, this norm is positive 
if $\tom =\om -m\Om_{\rm H} > 0$. When $\tom < 0$, the norm could be either 
positive or negative depending on the behavior of the solution in 
$r,\, \theta$. If the norm of $u_{\om lm}(x)$ is negative, we can easily 
see that $u_{-\om l-m}(x)$ has the positive norm. Let us define a set
$N^{-}$ consisting of mode solutions $u_{\om lm}$ with $\om >0$ whose norms 
are negative. Then, after suitable normalizations, these real frequency 
modes will satisfy the following orthogonality relations:
\beqa
<\! u_{\om lm}\, ,\, u_{\om^{\pr}l^{\pr}m^{\pr}}\! > = 
\delta (\om -\om^{\pr})\delta_{ll^{\pr}}\delta_{mm^{\pr}}   \qquad
{\rm for} \qquad  u_{\om lm} \not\in  N^{-},   \nonumber    \\
<\! u_{-\om l-m}\, ,\, u_{-\om^{\pr}l^{\pr}-m^{\pr}}\! > = 
\delta (\om -\om^{\pr})\delta_{ll^{\pr}}\delta_{mm^{\pr}}   \qquad
{\rm for} \qquad u_{\om lm} \in  N^{-}.
\label{OrthoR}
\eeqa
This set of solutions forms a complete basis of real frequency normal 
mode solutions having positive norms.

For complex frequency normal mode solutions, the property of inner 
products is very different. As already mentioned, the norm of 
\beq
v_{\om lm}(x) 
=\f{v_{\om lm}(r)}{\sq{r^2+a^2}}S_{\om lm}(\theta )e^{-i\om t+im\varphi}
\eeq
is zero. However, the inner product between $v_{\om lm}(x)$ and 
$v_{\om^{\ast}lm}(x)$ is nonzero
\beq
<\! v_{\om lm}\, ,\, v_{\om^{\ast}lm}\! > = 
\int (\om^{\ast}-m\Om )[\f{v^{\ast}_{\om lm}(r)}{\sq{r^2+a^2}}
S^{\ast}_{\om lm}(\theta )]^2 N^{-1}d\Sigma  .
\eeq
After suitably normalizing $v_{\om lm}(r)$, we can set it to be unit.
Then,
\beq
<\! v^{\ast}_{\om lm}\, ,\, v^{\ast}_{\om^{\ast}lm}\! > =
- <\! v_{\om lm}\, ,\, v_{\om^{\ast}lm}\! >^{\ast} = -1.
\label{OrthoC}
\eeq
All other inner products vanish. As mentioned above, by linearly combining 
$v_{\om lm}$ and $v_{\om^{\ast}lm}$, for example, one can construct 
a solution whose associated classical energy defined in Eq.~(\ref{CE})
is negative.  

Finally, it should be pointed out that any normal mode solution with
complex frequency can not be expressed by linearly combining the real
frequency normal modes $\{ u_{\om lm}(x)\} $. 
It follows because otherwise the complex frequency mode would not have 
net ingoing or outgoing flux.  
Therefore, the set of complex frequency normal mode solutions 
represents new independent degrees of freedom of the system, which can
describe field solutions carrying arbitrary values of energy including 
negative ones by linear combinations.

\section{Quantization}
\label{Q}

Based on the analysis of normal mode solutions for the classical scalar 
field in the previous section, we now proceed the canonical quantization
by interpreting the field $\phi (x)$ as an operator-valued distribution. 
The neutral scalar field can be expanded in terms of normal mode 
solutions as follows
\beqa
\phi (x) &=& \sum_{lm}\int_{\not\in N^{-}} d\om \f{1}{\sq{2}}
	     [a_{\la}u_{\la}(x)+a^{\dagger}_{\la}u^{\ast}_{\la}(x)]
	     +\sum_{lm}\int_{\in N^{-}} d\om \f{1}{\sq{2}}
	     [a_{-\la}u_{-\la}(x)+a^{\dagger}_{-\la}u^{\ast}_{-\la}(x)]
	     \nonumber   \\
	 & & \sum_{\om lm} \f{1}{\sq{2}}
	     [b_{\la}v_{\la}(x)+b^{\dagger}_{\la}v^{\ast}_{\la}(x)
	      +b_{\bl}v_{\bl}(x)+b^{\da}_{\bl}v^{\ast}_{\bl}(x)],
\label{Field}
\eeqa
where $\la$ denotes to $(\om ,l,m)$, $-\la$ to $(-\om ,l,-m)$, and 
$\bl$ to $(\om^{\ast},l,m)$. The expansion coefficients are now operators. 
We assume the equal-time commutation relations for $\phi (x)$ and 
$\pi (x)$
\beq
\lbrack \phi (x)\, ,\, \pi (y)\rbrack = i\delta^{(3)}(x, y)\, , \qquad 
\lbrack \phi (x)\, ,\, \phi (y) \rbrack  = 
\lbrack \pi (x)\, ,\, \pi (y)\rbrack  =0
\eeq
at a $x^0=y^0=t= {\rm const.}$ spacelike hypersurface. $\delta^{(3)}(x, y)$
is the three dimensional Dirac $\delta$ function defined by
\beq
\int \delta^{(3)}(x, y) f(y) \sq{-g} d^3y = f(x)
\eeq
at the $x^0=y^0=t= {\rm const.}$ surface. By using the following relations
\beqa
a_{\la} &=& \sq{2}<\! u_{\la}\, ,\, \phi (x)\! >, \qquad  
a^{\da}_{\la} =-\sq{2}<\! u^{\ast }_{\la}\, ,\, \phi (x)\! >, \qquad  
b_{\la}=\sq{2}<\! v_{\bl}\, ,\, \phi (x)\! >,  \nonumber  \\
b^{\da}_{\la} &=& -\sq{2}<\! v^{\ast}_{\bl}\, ,\, \phi (x)\! >, \qquad
b_{\bl}=\sq{2}<\! v_{\la}\, ,\, \phi (x)\! >,  \qquad 
b^{\da}_{\bl}=-\sq{2}<\! v^{\ast}_{\la}\, ,\, \phi (x)\! >,
\eeqa
we find commutation relations among mode operators
\beq
\lbrack a_{\la}\, ,\, a^{\dagger }_{\la^{\pr}} \rbrack =
\delta_{\la \la^{\pr}}~, \qquad 
\lbrack b_{\la}\, ,\, b^{\dagger }_{\bar{\la^{\pr}}} \rbrack =
\delta_{\la \la^{\pr}}~, \qquad    
\lbrack b_{\la}\, ,\, b^{\dagger }_{\la^{\pr}} \rbrack =
\lbrack b_{\bl}\, ,\, b^{\dagger }_{\bar{\la^{\pr}}} \rbrack =
\lbrack b_{\la}\, ,\, b_{\bar{\la^{\pr}}} \rbrack =0 ~.  
\eeq
All others vanish. Note that the real frequency mode operators satisfy
the usual commutation relations whereas mode operators for complex 
frequencies have unusual commutation relations as in other models 
in the Minkowski flat spacetime in Refs.~\cite{Schroer,SS,Fulling,Kang}.
In particular, $b_{\la}$ does commute with $b^{\da}_{\la}$. 

Now the Hamiltonian operator can be expressed in terms of mode 
operators by using Eq.~(\ref{Ham}) \cite{footnote2}:
\beqa
H &=& \f{1}{2} \sum_{lm}\int_{\not\in N^{-}} d\om \om (a^{\da}_{\la}
      a_{\la}+a_{\la}a^{\da}_{\la}) 
      + \f{1}{2} \sum_{lm}\int_{\in N^{-}} d\om (-\om )(a^{\da}_{-\la}
      a_{-\la}+a_{-\la}a^{\da}_{-\la})   \nonumber    \\
  & & +\f{1}{2}\sum_{\om lm}[\om (b_{\la}b^{\da}_{\bl}+b^{\da}_{\bl}
      b_{\la}) + \om^{\ast} (b^{\da}_{\la}b_{\bl}+b_{\bl}b^{\da}_{\la})],
\label{HamOp1}
\eeqa
where $\om >0$ for real frequency modes and $\om^{\rm I}>0$ for complex
frequency modes. Note first that $H$ is Hermitian, $H^{\da}=H$, as 
expected. For real frequency modes, let
\beq
a_{\pm \la}=\f{1}{\sq{2}}(\sq{\om}Q_{\pm \la}+i\f{P_{\pm \la}}{\sq{\om}}),
\eeq
where $Q_{\pm \la}$ and $P_{\pm \la}$ are Hermitian operators satisfying
$\lbrack Q_{\pm \la}\, ,\, P_{\pm \la} \rbrack = i$. Then one can easily 
see that the Hamiltonian for real frequency modes has a representation 
of a set of {\it attractive} harmonic oscillators as usual. 
Thus its energy spectrum is discrete. Interestingly, however, the energy
associated with the second term in Eq.~(\ref{HamOp1}) is always 
{\it negative} and {\it bounded above} whereas the first term shows 
positive and lower bounded energy. Due to this, although a vacuum state
can be defined such that $a_{\pm \la}|0\! >_{\!\! \rm R}=0$
for all $\la$, it is not the state of the lowest energy and in fact
there is no such state. All energy eigenstates can be constructed from
$|0\! >_{\!\! \rm R}$ simply by $(a^{\da}_{\pm \la})^n
|0\! >_{\!\! \rm R}$. Therefore, real frequency mode operators 
possess the usual symmetrized Fock representation ${\cal H}^{\rm R}$
as well as the particle interpretation.

For complex frequency modes, let 
\beq
H^C_{\la} =\f{1}{2}[\om (b_{\la}b^{\da}_{\bl}+b^{\da}_{\bl}
      b_{\la}) + \om^{\ast} (b^{\da}_{\la}b_{\bl}+b_{\bl}b^{\da}_{\la})].
\eeq
Suppose that it has an energy eigenstate $|E\!\! >$ such that $H^C_{\la}
|E\!\! > = E|E\!\! >$. Note then that
\beq
\lbrack H^C_{\la}, b^{\da}_{\bl} \rbrack =\omega b^{\da}_{\bl}, \qquad
\lbrack H^C_{\la}, b_{\la} \rbrack =-\omega b_{\la}, \qquad
\lbrack H^C_{\la}, b^{\da}_{\la} \rbrack =\om^{\ast} b^{\da}_{\la}, \qquad
\lbrack H^C_{\la}, b_{\bl} \rbrack =-\om^{\ast} b_{\bl}.
\eeq
Thus $b^{\da}_{\bl}|E\!\! >$, $b_{\la}|E\!\! >$ and $b^{\da}_{\la}|E\!\! >$,
$b_{\bl}|E\!\! >$ are eigenstates with eigenvalues of $(E\pm \om )$ 
and $(E\pm \om^{\ast})$, respectively, which are no longer real.
Since $H^C_{\la}$ is Hermitian, it presumably implies that energy eigenstates 
are not normalizable. This property shall be explicitly shown below.

To find the energy spectrum and its representation for $H^C_{\la}$, let 
us use some methods developed in Refs.~\cite{Fulling,SS,Schroer}. 
Let 
\beqa
b_{\la} &=& \f{1}{2}[i(\sq{\om^{\rm I}} q_{1\la}+\f{1}{\sq{\om^{\rm I}}}
	    p_{1\la}) + (\sq{\om^{\rm I}}q_{2\la}+\f{1}{\sq{\om^{\rm I}}}
	    p_{2\la})],    \nonumber   \\
b^{\da}_{\bl} &=& \f{1}{2}[(\sq{\om^{\rm I}} q_{1\la}-\f{1}{\sq{\om^{\rm I}}}
	    p_{1\la}) + i(\sq{\om^{\rm I}}q_{2\la}-\f{1}{\sq{\om^{\rm I}}}
	    p_{2\la})].
\eeqa
Here $q$ and $p$ are Hermitian operators satisfying 
$\lbrack q_{j\la}, p_{j\la} \rbrack = i$, $j=1,\, 2$. We find then 
\beq
H^C_{\la} = \f{1}{2}(p^2_{1\la}-(\om^{\rm I})^2q^2_{1\la})
            +\f{1}{2}(p^2_{2\la}-(\om^{\rm I})^2q^2_{2\la})
	    +\om^{\rm R}(q_{1\la}p_{2\la}-p_{1\la}q_{2\la}).
\eeq
Classical equations of motion corresponding to this Hamiltonian will be
\beq
\ddot{q_1}=|\om |^2q_1-2\om^{\rm R}\dot{q_2} \, , \qquad 
\ddot{q_2}=|\om |^2q_2+2\om^{\rm R}\dot{q_1}.
\eeq
Thus this is a system of two coupled {\it inverted} harmonic oscillators
with the same frequency $|\om |$. 

As in Ref.~\cite{SS}, the Hamiltonian operator $H^C_{\la}$ can be realized
as a sum of the infinitesimal generators of dilatations and rotations
acting on a function space $L^2({\bf R}^2)$ in a two-dimensional
Euclidean space \cite{footnote3}. Energy eigenfunctions are 
\beq
\psi_{\ve k}(\rho , \phi )= (2\pi )^{-1}\rho^{i\ve -1}e^{ik\phi},
\eeq
where $\ve$ is any continuous real number and $k$ any integer. 
$(\rho, \phi )$ are the polar coordinates. These eigenfunctions are
orthogonal
\beq
\int \psi^{\ast}_{\ve^{\pr}k^{\pr}}({\bf x})\psi_{\ve k}({\bf x})
     \rho d\rho d\phi = \delta (\ve -\ve^{\pr})\delta_{kk^{\pr}},
\label{Ortho}
\eeq
and form a complete set
\beq
\int^{\infty}_{-\infty} \psi^{\ast}_{\ve k}({\bf x})\psi_{\ve k}
({\bf y}) d\ve = \delta^{(2)}({\bf x}-{\bf y}).
\label{Com}
\eeq
The energy spectrum for $H^C_{\la}$ is
\beq
E_{\ve_{\la}k_{\la}}=\om^{\rm I}\ve_{\la}+\om^{\rm R}k_{\la}.
\eeq
It shows that
the energy eigenvalue is {\it continuous} for given $k_{\la}$ 
and {\it unbounded} below. Eq.~(\ref{Ortho}) shows explicitly that 
energy eigenstates are not normalizable as we expected above. However, 
we can construct normalizable wave packets from them. These square
integrable wave packets, for example, $|\psi \!\! >=\sum_{k_{\la}}
\int d\ve_{\la}<\!\! \psi_{\ve_{\la}k_{\la}}| \psi \!\! > 
|\psi_{\ve_{\la}k_{\la}}\!\!>$, will form a Hilbert space 
${\cal H}^C_{\la}$ which is isomorphic to $L^2({\bf R}^2)$. 

Any quantum state of the field which is in this Hilbert space will give 
rise to instability. It follows because, although the total energy of this
state is definite and time independent, the energy density outside the 
ergoregion will be positive and have exponential time dependence 
whereas the energy density within the ergoregion will have the same 
behavior but with negative energy, keeping the total energy over
the whole space fixed. Therefore, an observer sitting outside 
the ergoregion will measure time dependent radiation of positive energy. 
In addition, since the energy spectrum is unbounded below, some external
interaction with this system can give energy extraction from the system
without bound. 

Finally, we complete our quantization of the field $\phi (x)$ 
by constructing the total Hilbert space as follows,
\beq
{\cal H} = {\cal H}^{\rm R} \otimes \prod_{\la} {\cal H}^C_{\la}.
\eeq
Here ${\cal H}^{\rm R}$ is the usual symmetrized Fock space generated
by real frequency modes and $\prod_{\la} {\cal H}^C_{\la}$ is the infinite
number of products of Hilbert spaces ${\cal H}^C_{\la}$ generated by
complex frequency modes.

\section{Quantum instability}
\label{QI}

In this section, let us look at some interesting properties of the 
Hilbert space ${\cal H} = {\cal H}^{\rm R} \otimes \prod_{\la} 
{\cal H}^C_{\la}$ constructed in the previous section. First of all,
one may ask whether or not this Hilbert space still possesses the
particle interpretation of the quantum field. For real frequency modes, 
the energy spectrum is discrete and a vacuum state is defined well in
the usual way. $a^{\da}_{\pm \la}$ and $a_{\pm \la}$ can still be 
interpreted as creation and annihilation operators of energy quanta of
$\pm \hbar \om$, respectively. Thus real frequency mode operators still
have particle interpretation as usual except that the vacuum state 
defined is not the lowest energy state any more. 

For complex frequency modes, however, since the corresponding energy 
spectrum is continuous, it would not be possible to define some energy 
quanta whose multiples cover the whole energy spectrum. Therefore mode
operators associated with complex frequencies do not have particle 
interpretation in the usual sense. However, we may expect that there
will be the energy quanta of $\hbar \om^{\rm R}$ since the spectrum
of the rotation generator $Q_{\la}= q_{1\la}p_{2\la}-p_{1\la}q_{2\la}$
is discrete \cite{footnote4}. 

Now let us see how the appearance of an ergoregion at the late stage of
a dynamically evolving background spacetime starts to give a spontaneous 
radiation of energy. We expect this spontaneous quantum radiation 
if the initial vacuum state $|0>_{\!\! \rm in}$ of the field 
in the past falls in  any state in $\prod_{\la} {\cal H}^C_{\la}$    
in the remote future. To see this effect let us consider 
a ``paticle"detector linearly coupled to the field near 
$t \sim \infty$ placed in the in-vacuum state 
$|0>_{\!\! \rm in}$. The transition probability of the detector is 
proportioal to the response function ${\cal F}(E)$ \cite{BD}.
\beq
{\cal F}(E)=\lim_{t_0\to \infty }\int^{t_0+T}_{t_0}dt 
\int^{t_0+T}_{t_0}dt'\, e^{-iE(t-t')} 
{_{\rm in}\!\! }<\! 0|\phi [x(t)]\phi [x(t')]|0\! >_{\!\! \rm in}. 
\eeq
The field operator can also be decomposed as follows
\beq
\phi (x) = \sum_{lm}\int dk \f{1}{\sq{2}}(c_{\s}U_{\s}(x)+c^{\da}_{\s}
	   U^{\ast}_{\s}),
\eeq	   
where $U_{\s}(x)=U_{klm}(x)$ becomes spherical waves in the flat 
spacetime in the past infinity, and $c_{\s}|0\! >_{\!\! \rm in}=0$ for all 
$\s =(k,l,m)$. Since ${_{\rm in}\!\! }<\! 0|\phi [x(t)]\phi [x(t')]
|0\! >_{\!\! \rm in}=\f{1}{2}\sum_{\s} U_{\s}(x)U^{\ast}_{\s}(x^{\pr})$,
we now see
\beq
{\cal F}(E)=\lim_{t_0\to \infty }\f{1}{2}\sum_{\s}\, 
|\int^{t_0+T}_{t_0}\! dt\, e^{-iEt}U_{\s}(x)\, |^2 .    
\label{Rate}
\eeq
Since $\{ U_{\s},\, U^{\ast}_{\s}\}$ consists of a complete set,
all out normal modes can be expressed by this set. Let the Bogolubov 
transformations be 
\beq   
u_{\la} = \sum_{\s}(\al_{\la \s}U_{\s}
	    + \beta_{\la \s}U^{\ast}_{\s}), \quad   
v_{\la} = \sum_{\s}(\gamma_{\la \s}U_{\s}
	    + \eta_{\la \s}U^{\ast}_{\s}), \quad    
v_{\bl} = \sum_{\s}(\gamma_{\bl \s}U_{\s}
	    + \eta_{\bl \s}U^{\ast}_{\s}).
\label{Bogol1}
\eeq   
From orthogonality relations among out modes in Eq.~(\ref{OrthoR}) and
Eq.~(\ref{OrthoC}), we have
\beq
\sum_{\s}(\al^{\ast}_{\la \s}\al_{\la^{\pr}\s}
-\beta^{\ast}_{\la \s}\beta_{\la^{\pr}\s})
=\delta_{\la \la^{\pr}},   \qquad   
\sum_{\s}(\al_{\la \s}\beta_{\la^{\pr}\s}
-\beta_{\la \s}\al_{\la^{\pr} \s}) =0 
\eeq
for real frequency modes, and
\beq   
\sum_{\s}(\gamma^{\ast}_{\la \s}
\gamma_{\la^{\pr}\s} - \eta^{\ast}_{\la \s}
\eta_{\la^{\pr}\s})
= \sum_{\s}(\gamma^{\ast}_{\bl \s}
\gamma_{\bar{\la^{\pr}}\s} - \eta^{\ast}_{\bl \s}
\eta_{\bar{\la^{\pr}}\s}) =0,   \quad    
\sum_{\s}(\gamma^{\ast}_{\la \s}
\gamma_{\bar{\la^{\pr}}\s} - \eta^{\ast}_{\la \s}
\eta_{\bar{\la^{\pr}}\s}) = \delta_{\la \la^{\pr}}
\label{BogolC}
\eeq   
for complex frequency modes. Note the ``converted" relations 
among Bogolubov coefficients for complex frequency modes 
which are resulted from the unusual form of the orthogonal relations 
for such modes. Equivalently, we have from Eq.~(\ref{Bogol1}) 
\beqa
U_{\s} &=& \sum_{lm}\int_{\not\in N^{-}} d\om 
	    (\al^{\ast}_{\la \s}u_{\la}-\beta_{\la \s}
	    u^{\ast}_{\la}) 
	    +\sum_{lm}\int_{\in N^{-}} d\om
	    (\al^{\ast}_{-\la \s}u_{-\la}-\beta_{-\la \s}
	    u^{\ast}_{-\la})  \nonumber  \\
& & +\sum_{\om lm}(\gamma^{\ast}_{\bar{\la}\s}
    v_{\la}-\eta_{\bar{\la}\s}v^{\ast}_{\la}
    +\gamma^{\ast}_{\la \s}v_{\bar{\la}} 
    -\eta_{\la \s}v^{\ast}_{\bar{\la}}).
\label{Bogol2}
\eeqa

The response function ${\cal F}(E)$ in Eq.~(\ref{Rate}) will depend on 
the polar angle $\theta$, but be independent on $\varphi$ 
because of the axial symmetry of the background spacetime.
For calculational simplicity, we consider the following response function 
integrated over $\theta$; $0 \sim \pi$, 
\beq
F(E) = 2\pi \int^{\pi}_{0} {\cal F}(E) \sin \theta d\theta .
\eeq
Now, if the detector is at rest near spatial infinity, we obtain, 
from the asymptotic behavior of normal modes in Eq.~(\ref{ASolR}) 
and Eq.~(\ref{ASolC}), the transition rate
\beqa
\f{F(E)}{T} &\sim & \f{1}{2}\sum_{\s} \{\sum_{lm}
	\int_{\not\in N^{-}} d\om |\beta_{\la \s}|^2
	|\f{u_{\la}(r)}{\sq{r^2+a^2}}|^2\delta (E-\om ) 
	\nonumber   \\
& & + \sum_{lm}\int_{\in N^{-}} d\om
	|\al_{-\la \s}|^2|\f{u_{-\la}(r)}{\sq{r^2+a^2}}|^2
	\delta (E-\om )    \nonumber   \\
& & - \sum_{\om lm}2{\rm Re}[\gamma^{\ast}_
      {\bar{\la}\s}
      \gamma_{\la \s}(\f{v_{\la}(r)}{\sq{r^2+a^2}})^2 
      \f{e^{-i(E+\om^{\rm R})T}}{(E+\om )^2}  \nonumber   \\
& & +\eta_{\bar{\la}\s}\eta^{\ast}_{\la \s}
      (\f{v^{\ast}_{\la}(r)}{\sq{r^2+a^2}})^2
      \f{e^{-i(E-\om^{\rm R})T}}{(E-\om^{\ast})^2}]
      \f{e^{\om^{\rm I}T}}{T} \}
\eeqa
for large $T \gg 1$. This result in general shows non-vanishing 
excitations of the particle detector related to complex frequency 
modes as well as the usual contributions due to the mode mixing in real 
frequency modes. In particular, the contributions related to complex 
frequency modes are not stationary, but exponentially increasing 
in time $T$ \cite{footnote5}. The $\delta$-function dependence 
in the first two terms implies the energy conservation; that is, 
only the real frequency mode whose quantum energy is the same as 
that of the particle detector $(\om^{\pr} =E)$ can excite the detector. 
For complex frequency modes, however, all modes contribute to 
the excitation possibly because the energy spectrum for any complex 
frequency mode is continuous.

The first two terms will in general appear because the background 
spacetime is evolving in time and so it will effectively give 
time-dependent potential in Eq.~(\ref{KGR2}). However, in the case 
that there is no mixing in positive real frequency modes, they will 
vanish. The last term will vanish in the case that both 
$\gamma^{\ast}_{\bar{\la}\s}\gamma_{\la \s}$ and 
$\eta_{\bar{\la}\s}\eta^{\ast}_{\la \s}$ 
vanish for {\it all} $\la$ and $\s$. However, we do not expect this case 
if there is any instability mode due to the presence of the ergoregion
since otherwise the last equality in Eq.~(\ref{BogolC}) cannot be 
satisfied.

\section{Discussion}
\label{D}

In this paper, we have shown how the canonical quantization 
for a scalar field can be formulated in the presence of unstable modes 
due to ergoregion in a certain spacetime model. 
We found that our quantization has essentially the same features 
as in other models in Minkowski flat spacetime 
in Refs.~\cite{Fulling,SS,Kang}. The Hamiltonian operator for complex 
frequency modes is equivalent to a system of a set of two coupled 
{\it inverted} harmonic oscillators. Thus the energy spectrum is 
continuous and unbounded below. Consequently, the corresponding 
Hilbert space is not a Fock-like representation and has no particle 
interpretation for such complex frequency mode operators. 
The ``particle" detector placed in the in-vacuum state shows that 
a rotating star with ergoregion but without horizon has the quantum 
instability as well, leading exponentially time dependent 
spontaneous energy radiation 
to spatial infinity. This quantum instability is possible because 
negative energy could be accumulated within the ergoregion.
Accordingly, our result resolves the contradiction between conclusions 
in Ref.~\cite{MDO} and Ref.~\cite{AM2}.

Although the spacetime model considered in this paper is physically 
plausible, it is not so realistic. Thus one may ask whether 
the essential result obtained here would remain in a more realistic 
model. In the asymptotic region at spatial infinity, there would be 
no much difference in the analysis. Near the rotating object and inside 
it, the forms of normal mode solutions will be quite different from
ours. As long as an ergoregion is present, however, there will still be
two classes of normal mode solutions, that is, one described by real
frequencies and the other by complex frequencies. Similar inner product
properties shown in Sec.~\ref{CF} will still hold for this set of normal
mode solutions since they do not depend on the details of the form of 
solutions. Then the rest of the quantization procedure and the structure
of the corresponding total Hilbert space will be same. Therefore, the 
main results in our model will remain in a more realistic spacetime
model as well. As pointed out in Ref.~\cite{MDO}, 
the mirror boundary condition that we assumed to avoid 
the difficulty of obtaining complicated solutions inside the rotating
body in fact mimics effectively the center of the star provided that 
the modes propagate freely through the interior without much interactions
with the star body. In the limiting case that $\tr_0 \rightarrow -\infty$, 
our spacetime model in the future infinity becomes the rotating Kerr 
black hole. In this case our results should agree with Hawking's for 
the case of collapsing rotating black holes \cite{Hawking}. 
From eq.~(\ref{Imaginary}), we see that the imaginary part of the complex
frequency becomes zero as $\tr_0 \rightarrow -\infty$, and, accordingly, 
only real frequency modes $\{ u_{\om lm}(x) \}$ appear in our model. 
However, it is as yet unclear whether or not this mode decomposition 
constructed here in the future infinity gives the same vacuum as
Hawking's \cite{footnote8}. 

It will be straightforward to extend our formalism to other matter fields 
such as massive charged scalar and electromagnetic fields. For spinor 
fields, however, it is unclear at the present whether or not rotating 
stars with ergoregions spontaneously radiate fermionic energy to
infinity as well. It is because spinor fields do not give superradiance
in the presence of an ergoregion. Hence we cannot apply the heuristic 
argument in Sec.~\ref{In} for the generation of instability modes. 
In fact, we find that the inner product defined in Ref.~\cite{Unruh} 
for spinor fields is still positive definite in our spcetime model. 
Then, as explained below Eq.~(\ref{OrthoG}), there exists no complex 
frequency mode and hence no unstable mode for spinor fields classically. 
However, as rotating black holes give fermion emissions in the quantum 
theory inspite of no superradiance at the classical level, there might 
be some quantum process through which the ergoregion gives fermionic
spontaneous energy radiation. In addition, the main result obtained 
in the algebraic approach \cite{AM2,AM1} does not seem to depend on 
which matter field is considered. 

As shown in preceding sections, the Hamiltonian operators associated 
with unstable modes do not admit a Fock-like representation, a vacuum 
state, or the particle interpretation of mode operators. Accordingly, 
the conventional analysis of the vacuum instability based on the uses 
of asymptotic vacua and appropriately defined number operators no longer 
applies to our case. However, we have shown that Unruh's ``particle" 
detector model, which indeed does not require the particle interpretation 
of the field, is still applicable for extracting some usefull physics 
in our case. In fact, our case serves as a good example illustrating 
the point of view that the fundamental object in quantum field theory
is the field operator itself, not the ``particles" defined in a preferred 
Fock space \cite{Field}. The expectation value of the energy-momentum
tensor operator, which is defined by field operators only, should also be 
a useful quantity in our case. To obtain meaningful expectation value, 
however, renormalization of the energy-momentum tensor would have to be 
understood first in the presence of such instability
modes \cite{footnote9}. As far as we know, this interesting issue has 
never been addressed in the literature. 

\begin{figure}[tbp]
\epsfysize=4.5cm
\hspace{5.5cm}
\epsfig{file=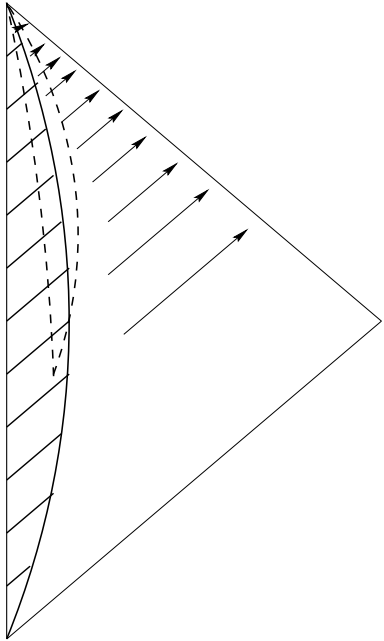,width=6cm,height=8.5cm}
\caption{Penrose diagram for a collapsing rotating object ending up to
         a stationary spacetime with an ergoregion.}
\label{Ergofig1}
\end{figure}

\begin{figure}[hbtp]
\epsfysize=4.5cm
\hspace{5.5cm}
\epsfig{file=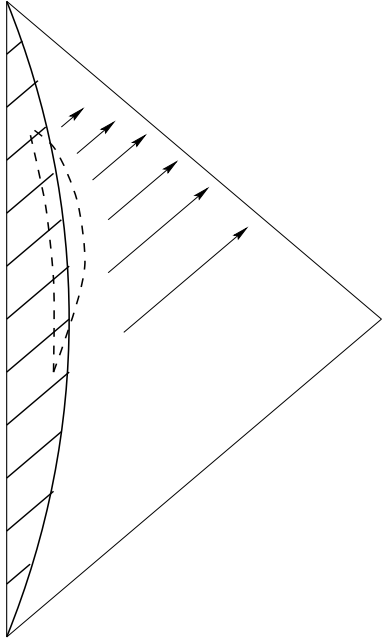,width=6cm,height=8.5cm}
\caption{Back reaction effect due to spontaneous evaporation of the
         ergosphere.}
\label{Ergofig2}
\end{figure}

The Penrose diagram for a realistic spacetime in which a stationary 
rotating star with an ergosphere develops in the remote future will be
as in Fig.~\ref{Ergofig1}. 
The solid line is the trajectory of the surface of a 
rotating object. The dotted lines denote the boundaries of the ergoregion.
Based on the analysis in our paper, exponentially time dependent 
spontaneous energy radiation will occur as soon as an ergoregion is formed. 
Then the back reaction of the quantum field on the metric will change
the gravitational fields of the evolving rotational object itself, 
depending on the strength and the time scale of the spontaneous radiation.
Since the wave trapped inside the ergoregion carries negative energy and 
the angular momentum in the opposite sense of the rotation, the rotating
object will loose its angular momentum and so the ergoregion can disappear
at some point of its evolution. Then the spontaneous radiation will also 
stops to occur. The corresponding Penrose diagram is shown 
in Fig.~\ref{Ergofig2}. 

It will also be very interesting to see how our quantization procedure in
this paper can be translated into algebraic approach. To obtain a 
quantum description in this approach, one defines first the $\ast$-algebra 
of the field operators and then constructs the Hilbert space of states 
by choosing an appropriate $\ast$-representation (equivalently, a suitable 
complex strucure) with a set of rules for dynamics. The most difficult 
part in this prescription is to single out the ``correct" representation 
among all possible $\ast$-representations. For spacetimes such as static or
stationary spacetimes \cite{footnote10}, certain physically motivated 
requirements select a unique complex structure and hence the ``correct" 
representation \cite{AM1}. For a spacetime where an ergoregion is 
present, on the other hand, Ashtekar and Magnon \cite{AM2} argue that 
there appears to exist no obvious way to choose even a specific complex
structure since ${\cal I}^-$ is not a Cauchy surface in this case 
\cite{footnote7}. However, the mode decomposition constructed 
in Sec.~\ref{CF} 
and in Sec.~\ref{QI} in our canonical quantization procedure suggests
that there may exist some way to construct the corresponding complex
structure in the algebraic approach as well. 

Finally, it should be pointed out that there are many other fields in 
physics in which complex frequency modes play important roles and so
our quantization formalism is potentially applicable. 
Generically, if a system stores some ``free"   
energy which can be released through interactions, then some amplifications
occur, revealing complex frequency modes classically. In a system of plasma,
for instance, a small perturbation of electric field exponentially 
increases in time if the phase velocity of the perturbed field is smaller
than the velocity of charged particles, and is damped in the opposite
case. The energy stored in plasma is released quickly by a small 
perturbation, giving complex frequency modes\cite{plasma}. In a tunable 
laser, the energy stored in dielectric material amplifies an incident
light and results in the intensity increment of the output laser beam.
In a field theoretic treatment of the system, the dielectric material
plays the role of a source producing an external potential and it is 
possible for complex frequency modes to occur under suitable conditions. 
In the theory of linear quantum amplifiers\cite{lamp}, one assumes a time 
dependent annihilation operator, $a(t)=a(0)e^{Wt/2-i\omega t}$ with a gain 
factor $W$. This gain factor $W$ may be interpreted simply as coming from 
the imaginary part of a complex frequency mode in the second quantization 
scheme where one does not need to assume the non-unitary evolution of the 
mode operator.
Therefore, the quantization formalism described at the present work may be 
useful to understand those phenomena in the context of quantum field theory.

\vskip 1cm
The author would like to thank for useful discussions with A. Ashtekar,  
D.R. Brill, B.R. Iyer, R. Nityananda, J. Samuel, M. Varadarajan,  
and J.H. Yee. Especially, I am deeply indebted to Ted Jacobson for many 
helpful discussions and suggestions. 



\begin{thebibliography}{99} 
\bibitem
{BI} E.M. Butterworth and J.R. Ipser, Ap. J. {\bf 204}, 200 (1975). 
\bibitem 
{SC} B.I. Schutz and N. Comins, Mon. Not. R. astr. Soc. {\bf 182}, 
     69 (1978).
\bibitem
{DI} S.L. Detweiler and J.R. Ipser, Astrophys. J. {\bf 185}, 675 (1973); 
     W.H. Press and S.A. Teukolsky, Astrophys. J. {\bf 185}, 649 (1973);
     S. Hawking, Commun. Math. Phys. {\bf 33}, 323 (1973). 
\bibitem
{Whiting} B.F. Whiting, J. Math. Phys. {\bf 30}, 1301 (1989).
\bibitem
{Staro} A.A. Starobinskii, Zh. Eksp. Teor. Fiz. {\bf 64}, 48 (1973)
	 [Sov. Phys. JETP {\bf 37}, 28 (1973)].
\bibitem
{Unruh} W.G. Unruh, Phys. Rev.D {\bf 10}, 3194 (1974).
\bibitem
{Ford} L.H. Ford, Phys. Rev. D {\bf 12}, 2963 (1975).
\bibitem
{AM1} A. Ashtekar and A. Magnon, Proc. R. Soc. Lond. A. {\bf 346}, 
      375 (1975). 
\bibitem
{Fried} J.L. Friedman, Commun. Math. Phys. {\bf 63}, 243 (1978).
\bibitem
{CS}  N. Comins and B.F. Schutz, Proc. R. Soc. Lond. A. {\bf 364},
      211 (1978). 
\bibitem
{Vilen} A. Vilenkin, Phys. Lett. {\bf 78B}, 301 (1978).
\bibitem
{AM2} A. Ashtekar and A. Magnon, C. R. Acad. Sci. Ser. A {\bf 281}, 
      875 (1975). 
\bibitem
{DKM} T. Dray, R. Kulkarni, and C.A. Manogue, Gen. Relativ. Gravit. 
      {\bf 24}, 1255 (1992). 
\bibitem
{MDO} A.L. Matacz, P.C.W. Davies and A.C. Ottewill, Phys. Rev. D {\bf 47},
      1557 (1993).    
\bibitem
{ZAMO} ZAMO observers are locally nonrotating observers whose trajectories 
       are tangent to $\pa_{\tau}=\pa_t+\Om \pa_{\varphi}$ where 
       $\Om =-\pa_t \cdot \pa_{\varphi}/{\pa_{\varphi}\cdot \pa_{\varphi}}$
       and $\pa_{\varphi}$ is the rotational Killing vector field of 
       stationary axisymmetric spacetimes. The trajectories of Killing
       observers are tangent to $\pa_t$ which becomes spacelike inside
       ergoregions. 
\bibitem
{SSW} L.I. Schiff, H. Snyder and J. Weinberg, Phys. Rev. {\bf 57}, 
      315 (1940). 
\bibitem
{SS} B. Schroer and J.A. Swieca, Phys. Rev. D {\bf 2}, 2938 (1970).
\bibitem
{Schroer} B. Schroer, Phys. Rev. D {\bf 3}, 1764 (1971). 
\bibitem
{Fulling} S.A. Fulling, {\it Aspects of Quantum Field Theory in Curved
	  Space-time}, (Cambridge University Press, Cambridge, England, 
	  1989).   
\bibitem
{Kang} G. Kang;``Quantization of scalar field in the presence of 
       imaginary frequency modes," UMDGR-96-036, RRI-96-13, 
       hep-th/9603166. 
\bibitem
{LB} L. Lindblom and D.R. Brill, Phys. Rev. D {\bf 10}, 3151 (1974); 
     see also J. Cohen, Phys.Rev. {\bf 173}, 1258 (1968) for stationary
     or adiabatically collapsing cases. 
\bibitem
{SEP} B. Carter, Commun. Math. Phys. {\bf 10}, 280 (1968); D. Brill,
      P.L. Chrzanowski, C.M. Pereira, E.D. Fackerell, and J.R. Ipser,
      Phys. Rev. D {\bf 5}, 1913 (1972). 
\bibitem
{footnote1} For the detailed functional form of $V_{\om lm}(r)$,
      see Ref.~\cite{Ford}. 
\bibitem
{footnote2} An easier way to compute the Hamiltonian operator is to use
      $\hat{\rm H}$ defined in Eq.~(\ref{FEQ}); ${\rm H}=<\! \Phi \, ,\, 
      \hat{\rm H}\Phi \! >$ where $\Phi$ is now the two-component 
      field operator. 
\bibitem
{footnote3} In Ref.~\cite{SS}, the dilatation operator $U(a)$ is defined 
      as $U(a)\psi (x, y)=e^{-a}\psi (e^{-a}x, e^{-a}y)$. 
      Our operators are expressed as;
      $q_{1\la}=(x+i\pa /{\pa x})/\sq{2\om^{\rm I}}$, $p_{1\la}=
      \sq{\om^{\rm I}/2}(x-i\pa /{\pa x})$, $q_{2\la}=(y+i\pa /{\pa y})
      /\sq{2\om^{\rm I}}$, and $p_{2\la}=\sq{\om^{\rm I}/2}
      (y-i\pa /{\pa y})$. We find then $H^C_{\la}=-i\om^{\rm I}
      (\rho \pa /{\pa \rho}+1)-i\om^{\rm R}\pa /{\pa \phi}$. The first
      term is the infinitesimal generator of dilatations and the second
      that of rotations. 
\bibitem
{footnote4} Note that $B^{\da}_{\la}=b^{\da}_{\la}b^{\da}_{\bl}$ and 
     $B_{\la}$ behave like creation and annihilation of energy quanta
     $2\hbar \om^{\rm R}$ (not $\hbar \om^{\rm R}$), respectively. 
\bibitem
{BD} N.D. Birrell and P.C.W. Davies, {\it Quantum Fields in Curved 
     Space} (Cambridge University Press, Cambridge, England, 1982).
\bibitem
{footnote5} Note, however, that it does not depend on the setting time
      $t_0$ of the detector. 
\bibitem
{Hawking} S.W. Hawking, Commun. Math. Phys. {\bf 43}, 199(1975).
\bibitem
{footnote8} The set $\{ u_{\om lm}(x) \}$ may not be equivalent to the mode
      set constructed in Ref.~\cite{Hawking} since $u_{\om lm}$ are solutions
      satisfying the mirror boundary condition. Work in progress. 
\bibitem
{Field} W.G. Unruh, ``Particles and Fields" in {\it Quantum Mechanics 
       in Curved Space-Time} edited by J. Audretsch and V. de Sabbata
       (Plenum Press, New York, 1990); R.M. Wald, {\it Quantum Field Theory 
       in Curved Spacetime and Black Hole Thermodynamics} (The University 
       of Chicago Press, Chicago, 1994).
\bibitem
{footnote9} Work in progress. 
\bibitem
{footnote10} By stationary spacetimes here we mean there exists a Killing 
       vector field $\xi_t =\partial_t$ which is timelike {\it everywhere}. 
\bibitem
{footnote7} The reason for this is that any data set on ${\cal I}^-$ can make 
      only positive contribution to the field energy and so such a data set
      can not recover a field with negative total energy which possibly 
      exists in presence of an ergoregion in the future \cite{AM2}. 
      However, this reasoning would not be valid in our case since the 
      spacetime is dynamically evolving and so is time dependent.  
\bibitem
{plasma} T.H. Stix, {\it The Theory of Plasma Waves} (McGraw-Hill 
      Book Company, New York, 1962).
\bibitem
{lamp} S. Stenholm, Physica Scripta, {\bf T12}, 56(1986).
%
\end{thebibliography}
\end{document}